\documentclass[a4paper,11pt]{article}
\usepackage{pos}
\usepackage{hyperref}
\newcommand{\as}{\alpha _s}
\newcommand{\eps}{\epsilon}
\newcommand{\mat}{\mathcal{M}}

\title{NNLOCAL: completely local subtractions}

\author*[a]{Flavio Guadagni}
\author[b]{Vittorio Del Duca}
\author[c]{Claude Duhr}
\author[d,e]{Levente Fek\'{e}sh\'{a}zy}
\author[d]{Pooja Mukherjee}
\author[f]{G\'{a}bor Somogyi}
\author[g]{Francesco Tramontano}
\author[f]{Sam Van Thurenhout}

\affiliation[a]{Physik Institut, Universit\"{a}t Z\"{u}rich,\\
  CH-8057 Zürich, Switzerland}
\affiliation[b]{INFN, Laboratori Nazionali di Frascati,\\
  00044 Frascati (RM), Italy}
\affiliation[c]{Bethe Center for Theoretical Physics, Universit\"{a}t Bonn, \\
D-53115, Germany}
\affiliation[d]{II. Institut f\"{u}r Theoretische Physik, Universität Hamburg,\\
  Luruper Chaussee 149, 22761, Hamburg, Germany}
\affiliation[e]{Institute for Theoretical Physics, ELTE E\"{o}tv\"{o}s Loránd University, \\
P\'{a}zm\'{a}ny P\'{e}ter s\'{e}t\'{a}ny 1/A, 1117, Budapest, Hungary}
\affiliation[f]{HUN-REN Wigner Research Centre for Physics, \\
Konkoly-Thege Mikl\'{o}s u. 29-33, 1121 Budapest, Hungary}
\affiliation[g]{Dipartimento di Fisica Ettore Pancini, \\
Universit\`{a} di Napoli Federico II and INFN - Sezione di Napoli, \\
Complesso Universitario di Monte Sant’Angelo Ed. 6, Via Cintia, 80126, Napoli, Italy}

\emailAdd{fguada@physik.uzh.ch}
\emailAdd{Vittorio.DelDuca@lnf.infn.it}
\emailAdd{cduhr@uni-bonn.de}
\emailAdd{levente.fekeshazy@desy.de}
\emailAdd{pooja.mukherjee@desy.de}
\emailAdd{somogyi.gabor@wigner.hun-ren.hu}
\emailAdd{francesco.tramontano@unina.it}
\emailAdd{sam.van.thurenhout@wigner.hun-ren.hu}

\abstract{The computation of higher-order corrections to cross sections relevant at LHC involves the 
evaluation of phase-space integrals that exhibit soft and collinear divergences. The subtraction of these 
divergences is a key ingredient to obtain fully-differential predictions for physical observables. We 
discuss a subtraction method to handle these divergences based on the construction of universal local 
counterterms. The integration of the counterterms is carried out analytically, giving a strong control on 
the numerical stability of our predictions. We implement our method in a numerical program, that we dub 
NNLOCAL, and validate it by computing the fully-differential NNLO cross-section for Higgs boson production 
in gluon-gluon fusion.}

\FullConference{The European Physical Society Conference on High Energy Physics (EPS-HEP2025)\\
7-11 July 2025\\
Marseille, France\\}

\begin{document}
\maketitle

\section{Introduction}

Our knowledge of elementary particles and their interaction is encoded in the Standard Model (SM) of particle physics. In the 
last decades, the SM has been intensively tested at collider experiments, like the Large Hadron Collider (LHC) at CERN, and the agreement among 
theoretical predictions and experimental data is remarkable. One of the most important confirmations of the SM was the 
discovery of the Higgs boson at LHC, in July 2012~\cite{ATLAS:2012yve,CMS:2012qbp}. Despite its successes, we know that the SM is not the end of the story. Some observed phenomena, 
like matter-antimatter asymmetry and the small neutrino masses, cannot be described within the SM framework. 
In addition to that, physics beyond the SM can manifest itself indirectly as a small deviation between 
SM predictions and experimental measurements. For all these reasons, precision tests of the SM become 
crucial.

At LHC, the effect of the strong force is predominant: for this reason it is necessary to rely on 
accurate Quantum Chromodynamics (QCD) calculations to make meaningful comparison with data. Theoretical 
predictions for observables at colliders are obtained using \textit{perturbation theory}: cross sections 
are expanded in a power series in the strong coupling constant $\as$. One of the key aspects for increasing 
precision is to include higher-order corrections in the perturbative expansion. This requires to take 
into account Feynman diagrams with additional real and virtual emissions with respect to the Born process. 
Both real and virtual diagrams suffer from infrared (IR) divergences, that are reached in the limits in which 
the partons are soft and/or collinear to each other. The sum of real and virtual contributions is finite, and 
therefore all the IR divergences cancel in the final result. However, the phase space integrations are typically not feasable in an analytical way, especially when considering differential distributions. For this reason, 
it is necessary to reorganize the various contributions of the computation such that the IR cancellation takes place explicitly, and a numerical integration can be carried out 
without additional complications. One possibility for doing so is offered by \textit{local subtraction methods.}
The key idea of local subtraction is to define some counterterms that have the same singular behaviour as 
the matrix element in the IR limits. The subtraction of these counterterms makes the phase space integration 
finite. However, to preserve the correct final result, they have to be added back integrated over the 
phase space of the additional radiation. The poles in the dimensional regulator $\epsilon $ of the integrated counterterms will cancel 
with the explicit poles of virtual amplitudes, leading to a finite expression suitable for numerical integration.

At Next-to-Leading Order (NLO), there are well established subtraction techniques~\cite{Frixione:1995ms,Catani:1996vz}, that have been implemented in several public codes. On the other side, the formulation of 
local subtraction methods at Next-to-Next-to-Leading order (NNLO) is still a very active field of research, and several approaches have been 
proposed~\cite{Gehrmann-DeRidder:2005btv,Czakon:2014oma,Caola:2017dug,Cacciari:2015jma,Magnea:2018hab,Herzog:2018ily}. 
Here, we focus on an extension of the CoLoRFulNNLO subtraction method~\cite{Somogyi:2005xz,Somogyi:2006db,Somogyi:2006da,DelDuca:2015zqa,DelDuca:2016csb,DelDuca:2016ily,Somogyi:2020mmk} to hadronic collisions. 
In the CoLoRFulNNLO method, the counterterms are defined based on the universal structure of QCD amplitudes in 
IR limits, that are extended over the whole phase space using suitable momentum mappings. The counterterms 
are then integrated analytically over the radiation phase space. We implement our method in a parton-level 
Monte Carlo code, that we dub \texttt{NNLOCAL}~\cite{DelDuca:2024ovc}, and as a proof-of-concept we compute the NNLO corrections to the cross 
section for Higgs boson production in gluon-gluon fusion in the Higgs effective field theory (HEFT) approximation. 

This proceeding is organised as follows: in Sec.~\ref{sec: subtraction}, we give a brief review of the 
CoLoRFulNNLO subtraction formalism. Sec.~\ref{sec: integration} is devoted to the analytic integration 
of the counterterms, while in Sec.~\ref{sec: nnlocal} we show some numerical results obtained with the \texttt{NNLOCAL} code. 
Finally, we draw our conclusions in Sec.~\ref{sec: conclusions}.

\section{The subtraction method}
\label{sec: subtraction}
We consider a hadron-hadron collision leading to the production of a colorless system $X$. At partonic level, 
the NNLO correction to the cross section for this process is 
\begin{equation}
\label{eq: NNLO cross section}
  \sigma ^{\rm NNLO}_{ab} = \int _{X+2} d\sigma ^{\rm RR}_{ab} J_{X+2} + \int _{X+1} \Big( d\sigma ^{\rm RV}_{ab} + d\sigma _{ab}^{C_1} \Big)  J_{X+1} + \int _X \Big( d\sigma ^{\rm VV}_{ab} + d\sigma_{ab}^{C_2}  \Big) J_X \, ,
\end{equation}
where $ d\sigma ^{\rm RR}_{ab}$, $ d\sigma ^{\rm RV}_{ab}$ and $ d\sigma ^{\rm VV}_{ab}$ are the double real, 
real-virtual and double virtual cross sections, while $ d\sigma ^{\rm C_1}_{ab}$ and $ d\sigma ^{\rm C_2}_{ab}$ 
are the collinear remnants. The function $J_X$ is a generic IR safe observable. In Eq.~(\ref{eq: NNLO cross section}), 
all the integrals are IR divergent, but their sum is finite. In order to regularize the IR singularities, we 
introduce several counterterms. Focusing on the double real contribution, it is regularized as 
\begin{equation}
\label{eq: regularized cross section}
  \sigma ^{\rm NNLO,RR}_{ab} = \int _{X+2} \Big[  d\sigma ^{\rm RR}_{ab} J_{X+2} -  d\sigma ^{\rm RR,A_1}_{ab}J_{X+1} -  d\sigma ^{\rm RR,A_2}_{ab}J_X +  d\sigma ^{\rm RR, A_{12}}_{ab}J_X \Big] \, ,
\end{equation}
that, by construction, it is finite in $d=4$ dimensions and can be integrated numerically. The counterterms 
introduced have the following interpretation: 
\begin{itemize}
  \item $d\sigma ^{\rm RR,A_1}_{ab}$ regularizes the singly-unresolved limits. 
  \item $d\sigma ^{\rm RR,A_2}_{ab}$ regularizes the doubly-unresolved limits. 
  \item $d\sigma ^{\rm RR,A_{12}}_{ab}$ cancels the singularity coming from the singly-unresolved limits of $d\sigma ^{\rm RR,A_2}_{ab}$, and from the doubly-unresolved limits of $d\sigma ^{\rm RR,A_1}_{ab}$.
\end{itemize}
The precise definitions of the counterterms are quite involved and we will not report them here. For the sake of 
this proceeding, it is necessary just to highlight some of their features. To cancel the IR singularities, the 
counterterms are defined such that they coincide with QCD matrix elements in the IR limits. Their extension over 
the whole phase space is done through suitable momentum mappings, that are defined such that the Born and 
the radiation phase spaces factorize. In addition to that, their definition is such that all the overlaps 
between singly and doubly unresolved limits are treated in a process-independent way.  
To preserve the result, each counterterm needs to be added back in its integrated version. In this proceeding we 
focus on the integration of the counterterm $d\sigma ^{\rm RR,A_2}_{ab}$ over the double-emission phase space: 
\begin{equation}
  \int _2 d\sigma ^{\rm RR,A_2}_{ab} \, .
\end{equation}

\section{Integration of the $A_2$ counterterm}
\label{sec: integration}

In this section we will discuss the computational strategy employed for the integration of the $d\sigma ^{\rm RR,A_2}_{ab}$ counterterm. It is defined to regularize the doubly-unresolved limits of matrix elements with two extra emissions, 
and its explicit definition is based on the double and triple collinear splitting kernels, and double soft 
eikonal factors~\cite{Catani:1999ss}. The analytical integration is performed by sorting all the terms appearing 
in the counterterm into \textit{integral topologies}, and after that we perform an Integration-By-Part (IBP)
reduction to a basis of \textit{master integrals} (MIs), using reverse unitarity~\cite{Anastasiou:2002yz} to construct IBP identities among phase space integrals. To cross-check our results, we compute the MIs using 
two different strategies: direct integration and the differential equations method~\cite{Kotikov:1991pm,Gehrmann:1999as}.

To illustrate our computational strategy, we consider an explicit example. The counterterm that regularizes the 
configurations in which the emitted partons $r$ and $s$ are collinear to the initial-state parton $a$ is 
\begin{align}
\label{eq: triple collinear counterterm}
  & \mathcal{C}^{IFF(0)}_{ars}(\{p\}_{X+2};p_a,p_b) = (8\pi \as \mu^{2\eps}) ^2 \frac{1}{x_{a,rs}}\frac{1}{s_{ars}^2} \nonumber \\
  & \times \langle \mat^{(0)}_{(ars)b,X}(\hat{p}_a,\hat{p}_b, \left\{ \hat{p} \right\}_{X}) | \hat{P}^{(0)}_{(ars)rs} | \mat^{(0)} _{(ars)b,X}(\hat{p}_a, \hat{p}_b, \{ \hat{p} \} _{X}) \rangle  \mathcal{F}(x_{a,rs},\xi_{a,rs}\xi_{b,rs})
\end{align}
where $\hat{P}^{(0)}_{(ars)rs} $ is the triple collinear splitting kernel~\cite{Catani:1999ss} and $s_{ij}=(p_i+p_j)^2$, where $i,j \in \{a,r,s\}$, and $ |\mat^{(0)}_{(ars)b,X}(\hat{p}_a,\hat{p}_b, \left\{ \hat{p} \right\}_{X})\rangle$ is the Born matrix element in color-spin space, evaluated on the set of mapped momenta $\{\hat{p}\}$. We point out that the mapped momenta depend on two convolution variables $\xi _a$ 
and $\xi _b$. 
The variables $x_{a,rs}$ and $\xi_{a,rs}, \xi_{b,rs}$ are defined as 
\begin{align}
  & x_{a,rs} = 1 - \frac{s_{ar} + s_{br}}{s_{ab}} - \frac{s_{as} + s_{bs}}{s_{ab}} \, , \nonumber \\
  & \xi _{a,rs} = \sqrt{\frac{s_{ab} - s_{b(rs)}}{s_{ab} - s_{a(rs)}} \frac{s_{ab} - s_{(rs)(ab)} + s_{rs}}{s_{ab}}} \, , \nonumber \\
  & \xi _{b,rs} = \sqrt{\frac{s_{ab} - s_{a(rs)}}{s_{ab} - s_{b(rs)}}\frac{s_{ab} - s_{(rs)(ab)} + s_{rs}}{s_{ab}} } \, ,
\end{align}
where we introduced the notations $s_{c(jk)} = s_{cj} + s_{ck}$ and $s_{(cd)(jk)} = s_{c(jk)} + s_{d(jk)}$. 
The function $\mathcal{F}(x,y)$ is defined as
\begin{equation}
  \mathcal{F}(x,y) = \left(\frac{x}{y} \right)^2 \, ,
\end{equation}
and its role is to cancel unphysical singularities associated to the factor $x_{a,rs}$.

The counterterm of Eq.~(\ref{eq: triple collinear counterterm}) has to be integrated over the double 
radiation phase space, whose integration measure is~\cite{Daleo:2006xa}
\begin{equation}
  d\Phi _{II,FF} = \frac{d^dp_r}{(2\pi)^{d-1}}\frac{d^dp_s}{(2\pi)^{d-1}} \delta _+(c_1)\delta _+(c_2)\delta _+(c_3)\delta _+(c_4) \, , 
\end{equation}
where the phase space constraints $c_1$ and $c_2$ are the on-shellness of the momenta $p_r$ and $p_s$, while 
the constraints $c_3$ and $c_4$ come from the phase space factorization. Explicitly, the constraints 
are given by
\begin{align}
  & c_1 = p_r^2, \quad c_2 = p_s ^2 \, , \nonumber \\
  & c_3 = (p_a + p_b -p_r -p_s)^2 - \xi _a \xi _b s_{ab} \, , \nonumber \\
  & c_4= \xi _a (s_{ab} - s_{ar} - s_{as}) + \xi _b (s_{ab} - s_{br} - s_{bs}) \, .
\end{align}
To derive IBP relations among phase space integrals, the delta functions are treated as cut propagators: 
\begin{equation}
  \delta _+ (c_i) \to \bigg(\frac{1}{c_i}\bigg)_c \, ,
\end{equation}
where in IBP relations, the cut propagators satisfy the condition
\begin{equation}
  \bigg(\frac{1}{c_i}\bigg)_c^\nu = 0 \, , \quad \text{for } \nu \leq 0 \, .
\end{equation}
We find that the integrated triple collinear counterterm is a linear combination of integrals 
belonging to 17 different integral topologies. For brevity, we will report only some of them, and 
the complete definition of the families will be given in a future publication
\begin{align}
  & F_1(n_1,n_2,n_3) = \int d\Phi _{II,FF} \frac{1}{s_{as}^{n_1}(s_{ar} + s_{as})^{n_2}(s_{ab}-s_{ar}-s_{br})^{n_3}} \, , \nonumber \\
  & F_2(n_1,n_2,n_3) = \int d\Phi _{II,FF} \frac{1}{s_{ar}^{n_1}(s_{ar} + s_{as})^{n_2} (s_{ar} + s_{br})^{n_3}} \, , \nonumber \\
  & F_3(n_1,n_2,n_3) = \int d\Phi _{II,FF} \frac{1}{s_{ar}^{n_1} (s_{ab} - s_{ar} - s_{br})^{n_2}(s_{ar} + s_{as} - s_{ab})^{n_3}} \, , \nonumber \\
  & F_4(n_1,n_2,n_3) = \int d\Phi _{II,FF} \frac{1}{s_{ar}^{n_1}(s_{ab} - s_{ar} - s_{br})^{n_2}(s_{ar} + s_{as} + s_{br} + s_{bs})^{n_3}} \, , \nonumber \\
  & \vdots \,  \nonumber \\
  & F_{17}(n_1,n_2,n_3)=\int d \Phi _{II,FF}\frac{1}{s_{as}^{n_1}(s_{ar}+s_{as}-s_{rs})^{n_2}(s_{ar}+s_{br})^{n_3}} \, . 
\end{align} 
After an IBP reduction we find that the integrated counterterm $\int _2 d\sigma ^{\rm RR, A_2}_{ab}$ 
is a linear combination of 42 master integrals. For each integral topology, we set up a system 
of differential equations in the convolution variables $\xi _a, \xi _b$, that is solved by finding a suitable transformation to a canonical 
basis~\cite{Henn:2013pwa}. After the phase space integration, the last step to be performed is 
the integration over the convolution variables. This is done by constructing a suitable 
distributional expansion. 

\section{The \texttt{NNLOCAL} code}
\label{sec: nnlocal}

The full subtraction formalism, including all the analytical expressions for the integrated 
counterterms, has been implemented in the parton level Monte Carlo code \texttt{NNLOCAL}~\cite{DelDuca:2024ovc}, that is publicly available at \url{https://github.com/nnlocal/nnlocal}. To 
test our code, we compute the inclusive NNLO cross section for Higgs boson production in gluon fusion in the Higgs effective field theory (HEFT) approximation. 
We find perfect agreement with the code \texttt{n3loxs}~\cite{Baglio:2022wzu}. Since \texttt{NNLOCAL} is 
fully differential in all the particle momenta, we also test it for the computation of differential observables. 
\begin{figure}[ht] 
  \centering
  \includegraphics[width=1.0\textwidth]{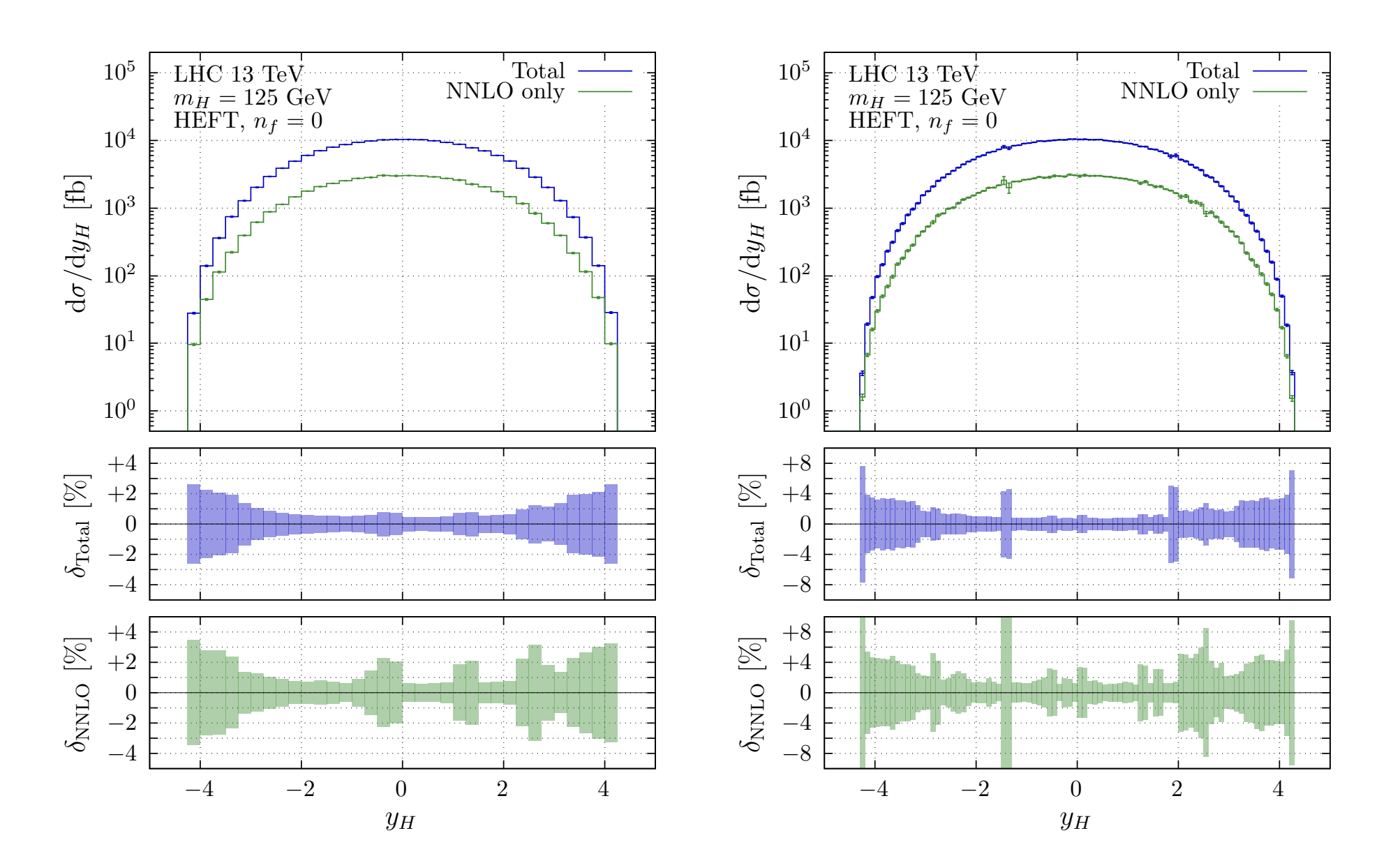} 
  \caption{The rapidity distribution of the Higgs boson at NNLO in the HEFT with $n_f=0$ light quarks. The distribution on the left has a bin width $\Delta y=0.25$, while the one on the right has a bin width of $\Delta y=0.1$. The lower panels show the relative error for the total distribution and the NNLO contribution. The error bands shown are the estimated Monte Carlo uncertanties.}
  \label{fig: rapidity distribution}
\end{figure}
In Fig.~\ref{fig: rapidity distribution} we present the rapidity distribution of a Higgs boson 
with mass $m_H=125$ GeV at the $13$ TeV at LHC in the HEFT approximation, considering $n_f=0$ light quarks. The runtime was 
about 1 hour on a MacBook Pro laptop with M2 processor and 8 CPU cores. In the left plot, with a 
bin width $\Delta y = 0.25$, we can see a good numerical convergence and stability. To test our code 
on more demanding conditions, we also plot the same distribution with a bin width of $\Delta y = 0.1$. 
We still see an overall good convergence, but because of the fine binning, some spikes arise due to the 
phenomenon of \textit{misbinning}. This problem can be solved by considering \textit{robust estimators} for the bin values. 

\section{Conclusions and outlook}
\label{sec: conclusions}

The computation of higher order corrections for collider observables is extremely important for precision tests of the Standard Model. The CoLoRFulNNLO method is designed to handle fully differential NNLO calculations numerically, by subtracting the IR divergences using suitable local 
counterterms. We perform an analytic integration of the counterterms needed for color singlet production 
in hadronic collisions at NNLO. The subtraction method is implemented in the public code \texttt{NNLOCAl}. 

Our proof-of-concept code is the first building block for the applications of the CoLoRFulNNLO method to hadronic collisions. Its future extension and refinement will make it a useful 
program to perform NNLO QCD computations for a variety of processes at LHC. 

\section*{Acknowledgments}

This work has been supported by grant K143451 of the National Research, Development and Innovation Fund in Hungary and the Bolyai Fellowship program of the Hungarian Academy of Sciences. The work of C.D.\ was funded by the European Union (ERC Consolidator Grant LoCoMotive 101043686). Views and opinions expressed are however those of the author(s) only and do not necessarily reflect those of the European Union or the European Research Council. Neither the European Union nor the granting authority can be held responsible for them. The work of L.F.\ was supported by the German Academic Exchange Service (DAAD) through its Bi-Nationally Supervised Scholarship program.

\bibliographystyle{unsrt}
\bibliography{biblio}

@article{ATLAS:2012yve,
    author = "Aad, Georges and others",
    collaboration = "ATLAS",
    title = "{Observation of a new particle in the search for the Standard Model Higgs boson with the ATLAS detector at the LHC}",
    eprint = "1207.7214",
    archivePrefix = "arXiv",
    primaryClass = "hep-ex",
    reportNumber = "CERN-PH-EP-2012-218",
    doi = "10.1016/j.physletb.2012.08.020",
    journal = "Phys. Lett. B",
    volume = "716",
    pages = "1--29",
    year = "2012"
}

@article{CMS:2012qbp,
    author = "Chatrchyan, Serguei and others",
    collaboration = "CMS",
    title = "{Observation of a New Boson at a Mass of 125 GeV with the CMS Experiment at the LHC}",
    eprint = "1207.7235",
    archivePrefix = "arXiv",
    primaryClass = "hep-ex",
    reportNumber = "CMS-HIG-12-028, CERN-PH-EP-2012-220",
    doi = "10.1016/j.physletb.2012.08.021",
    journal = "Phys. Lett. B",
    volume = "716",
    pages = "30--61",
    year = "2012"
}

@article{Frixione:1995ms,
    author = "Frixione, S. and Kunszt, Z. and Signer, A.",
    title = "{Three jet cross-sections to next-to-leading order}",
    eprint = "hep-ph/9512328",
    archivePrefix = "arXiv",
    reportNumber = "SLAC-PUB-7073, SLAC-PUB-95-7073, ETH-TH-95-42",
    doi = "10.1016/0550-3213(96)00110-1",
    journal = "Nucl. Phys. B",
    volume = "467",
    pages = "399--442",
    year = "1996"
}

@article{Catani:1996vz,
    author = "Catani, S. and Seymour, M. H.",
    title = "{A General algorithm for calculating jet cross-sections in NLO QCD}",
    eprint = "hep-ph/9605323",
    archivePrefix = "arXiv",
    reportNumber = "CERN-TH-96-029, CERN-TH-96-29",
    doi = "10.1016/S0550-3213(96)00589-5",
    journal = "Nucl. Phys. B",
    volume = "485",
    pages = "291--419",
    year = "1997",
    note = "[Erratum: Nucl.Phys.B 510, 503--504 (1998)]"
}

@article{Gehrmann-DeRidder:2005btv,
    author = "Gehrmann-De Ridder, A. and Gehrmann, T. and Glover, E. W. Nigel",
    title = "{Antenna subtraction at NNLO}",
    eprint = "hep-ph/0505111",
    archivePrefix = "arXiv",
    reportNumber = "ZU-TH-07-05, IPPP-05-18",
    doi = "10.1088/1126-6708/2005/09/056",
    journal = "JHEP",
    volume = "09",
    pages = "056",
    year = "2005"
}

@article{Czakon:2014oma,
    author = "Czakon, M. and Heymes, D.",
    title = "{Four-dimensional formulation of the sector-improved residue subtraction scheme}",
    eprint = "1408.2500",
    archivePrefix = "arXiv",
    primaryClass = "hep-ph",
    reportNumber = "TTK-14-16",
    doi = "10.1016/j.nuclphysb.2014.11.006",
    journal = "Nucl. Phys. B",
    volume = "890",
    pages = "152--227",
    year = "2014"
}

@article{Caola:2017dug,
    author = {Caola, Fabrizio and Melnikov, Kirill and R{\"o}ntsch, Raoul},
    title = "{Nested soft-collinear subtractions in NNLO QCD computations}",
    eprint = "1702.01352",
    archivePrefix = "arXiv",
    primaryClass = "hep-ph",
    reportNumber = "CERN-TH-2017-029, IPPP-17-10, TTP17-003",
    doi = "10.1140/epjc/s10052-017-4774-0",
    journal = "Eur. Phys. J. C",
    volume = "77",
    number = "4",
    pages = "248",
    year = "2017"
}

@article{Cacciari:2015jma,
    author = "Cacciari, Matteo and Dreyer, Fr{\'e}d{\'e}ric A. and Karlberg, Alexander and Salam, Gavin P. and Zanderighi, Giulia",
    title = "{Fully Differential Vector-Boson-Fusion Higgs Production at Next-to-Next-to-Leading Order}",
    eprint = "1506.02660",
    archivePrefix = "arXiv",
    primaryClass = "hep-ph",
    reportNumber = "CERN-PH-TH-2015-127, OUTP-15-12P",
    doi = "10.1103/PhysRevLett.115.082002",
    journal = "Phys. Rev. Lett.",
    volume = "115",
    number = "8",
    pages = "082002",
    year = "2015",
    note = "[Erratum: Phys.Rev.Lett. 120, 139901 (2018)]"
}

@article{Magnea:2018hab,
    author = "Magnea, L. and Maina, E. and Pelliccioli, G. and Signorile-Signorile, C. and Torrielli, P. and Uccirati, S.",
    title = "{Local analytic sector subtraction at NNLO}",
    eprint = "1806.09570",
    archivePrefix = "arXiv",
    primaryClass = "hep-ph",
    doi = "10.1007/JHEP12(2018)107",
    journal = "JHEP",
    volume = "12",
    pages = "107",
    year = "2018",
    note = "[Erratum: JHEP 06, 013 (2019)]"
}

@article{Herzog:2018ily,
    author = "Herzog, Franz",
    title = "{Geometric IR subtraction for final state real radiation}",
    eprint = "1804.07949",
    archivePrefix = "arXiv",
    primaryClass = "hep-ph",
    doi = "10.1007/JHEP08(2018)006",
    journal = "JHEP",
    volume = "08",
    pages = "006",
    year = "2018"
}

@article{Somogyi:2005xz,
    author = "Somogyi, Gabor and Trocsanyi, Zoltan and Del Duca, Vittorio",
    title = "{Matching of singly- and doubly-unresolved limits of tree-level QCD squared matrix elements}",
    eprint = "hep-ph/0502226",
    archivePrefix = "arXiv",
    reportNumber = "DFTT-05-05",
    doi = "10.1088/1126-6708/2005/06/024",
    journal = "JHEP",
    volume = "06",
    pages = "024",
    year = "2005"
}

@article{Somogyi:2006db,
    author = "Somogyi, Gabor and Trocsanyi, Zoltan",
    title = "{A Subtraction scheme for computing QCD jet cross sections at NNLO: Regularization of real-virtual emission}",
    eprint = "hep-ph/0609043",
    archivePrefix = "arXiv",
    doi = "10.1088/1126-6708/2007/01/052",
    journal = "JHEP",
    volume = "01",
    pages = "052",
    year = "2007"
}

@article{Somogyi:2006da,
    author = "Somogyi, Gabor and Trocsanyi, Zoltan and Del Duca, Vittorio",
    title = "{A Subtraction scheme for computing QCD jet cross sections at NNLO: Regularization of doubly-real emissions}",
    eprint = "hep-ph/0609042",
    archivePrefix = "arXiv",
    reportNumber = "DFTT-15-2006",
    doi = "10.1088/1126-6708/2007/01/070",
    journal = "JHEP",
    volume = "01",
    pages = "070",
    year = "2007"
}

@article{DelDuca:2015zqa,
    author = "Del Duca, Vittorio and Duhr, Claude and Somogyi, G{\'a}bor and Tramontano, Francesco and Tr{\'o}cs{\'a}nyi, Zolt{\'a}n",
    title = "{Higgs boson decay into b-quarks at NNLO accuracy}",
    eprint = "1501.07226",
    archivePrefix = "arXiv",
    primaryClass = "hep-ph",
    reportNumber = "CERN-PH-TH-2015-005, CP3-14-82",
    doi = "10.1007/JHEP04(2015)036",
    journal = "JHEP",
    volume = "04",
    pages = "036",
    year = "2015"
}

@article{DelDuca:2016csb,
    author = "Del Duca, Vittorio and Duhr, Claude and Kardos, Adam and Somogyi, G{\'a}bor and Tr{\'o}cs{\'a}nyi, Zolt{\'a}n",
    title = "{Three-Jet Production in Electron-Positron Collisions at Next-to-Next-to-Leading Order Accuracy}",
    eprint = "1603.08927",
    archivePrefix = "arXiv",
    primaryClass = "hep-ph",
    doi = "10.1103/PhysRevLett.117.152004",
    journal = "Phys. Rev. Lett.",
    volume = "117",
    number = "15",
    pages = "152004",
    year = "2016"
}

@article{DelDuca:2016ily,
    author = "Del Duca, Vittorio and Duhr, Claude and Kardos, Adam and Somogyi, G{\'a}bor and Sz{\H{o}}r, Zolt{\'a}n and Tr{\'o}cs{\'a}nyi, Zolt{\'a}n and Tulip{\'a}nt, Zolt{\'a}n",
    title = "{Jet production in the CoLoRFulNNLO method: event shapes in electron-positron collisions}",
    eprint = "1606.03453",
    archivePrefix = "arXiv",
    primaryClass = "hep-ph",
    reportNumber = "CERN-TH-2016-138, CP3-16-29, NSF-KITP-16-084",
    doi = "10.1103/PhysRevD.94.074019",
    journal = "Phys. Rev. D",
    volume = "94",
    number = "7",
    pages = "074019",
    year = "2016"
}

@article{Somogyi:2020mmk,
    author = "Somogyi, G{\'a}bor and Tramontano, Francesco",
    title = "{Fully exclusive heavy quark-antiquark pair production from a colourless initial state at NNLO in QCD}",
    eprint = "2007.15015",
    archivePrefix = "arXiv",
    primaryClass = "hep-ph",
    doi = "10.1007/JHEP11(2020)142",
    journal = "JHEP",
    volume = "11",
    pages = "142",
    year = "2020"
}

@article{Catani:1999ss,
    author = "Catani, Stefano and Grazzini, Massimiliano",
    title = "{Infrared factorization of tree level QCD amplitudes at the next-to-next-to-leading order and beyond}",
    eprint = "hep-ph/9908523",
    archivePrefix = "arXiv",
    reportNumber = "CERN-TH-99-263, ETH-TH-99-22",
    doi = "10.1016/S0550-3213(99)00778-6",
    journal = "Nucl. Phys. B",
    volume = "570",
    pages = "287--325",
    year = "2000"
}

@article{Anastasiou:2002yz,
    author = "Anastasiou, Charalampos and Melnikov, Kirill",
    title = "{Higgs boson production at hadron colliders in NNLO QCD}",
    eprint = "hep-ph/0207004",
    archivePrefix = "arXiv",
    reportNumber = "SLAC-PUB-9273",
    doi = "10.1016/S0550-3213(02)00837-4",
    journal = "Nucl. Phys. B",
    volume = "646",
    pages = "220--256",
    year = "2002"
}

@article{Kotikov:1991pm,
    author = "Kotikov, A. V.",
    title = "{Differential equation method: The Calculation of N point Feynman diagrams}",
    doi = "10.1016/0370-2693(91)90536-Y",
    journal = "Phys. Lett. B",
    volume = "267",
    pages = "123--127",
    year = "1991",
    note = "[Erratum: Phys.Lett.B 295, 409--409 (1992)]"
}

@article{Gehrmann:1999as,
    author = "Gehrmann, T. and Remiddi, E.",
    title = "{Differential equations for two-loop four-point functions}",
    eprint = "hep-ph/9912329",
    archivePrefix = "arXiv",
    reportNumber = "TTP-99-49",
    doi = "10.1016/S0550-3213(00)00223-6",
    journal = "Nucl. Phys. B",
    volume = "580",
    pages = "485--518",
    year = "2000"
}

@article{DelDuca:2024ovc,
    author = "Del Duca, V. and Duhr, C. and Fekeshazy, L. and Guadagni, F. and Mukherjee, P. and Somogyi, G. and Tramontano, F. and Van Thurenhout, S.",
    title = "{NNLOCAL: completely local subtractions for color-singlet production in hadron collisions}",
    eprint = "2412.21028",
    archivePrefix = "arXiv",
    primaryClass = "hep-ph",
    reportNumber = "DESY-24-211, BONN-TH-2024-18",
    doi = "10.1007/JHEP05(2025)151",
    journal = "JHEP",
    volume = "05",
    pages = "151",
    year = "2025"
}

@article{Henn:2013pwa,
    author = "Henn, Johannes M.",
    title = "{Multiloop integrals in dimensional regularization made simple}",
    eprint = "1304.1806",
    archivePrefix = "arXiv",
    primaryClass = "hep-th",
    doi = "10.1103/PhysRevLett.110.251601",
    journal = "Phys. Rev. Lett.",
    volume = "110",
    pages = "251601",
    year = "2013"
}

@article{Baglio:2022wzu,
    author = "Baglio, Julien and Duhr, Claude and Mistlberger, Bernhard and Szafron, Robert",
    title = "{Inclusive production cross sections at N$^{3}$LO}",
    eprint = "2209.06138",
    archivePrefix = "arXiv",
    primaryClass = "hep-ph",
    reportNumber = "CERN-TH-2022-109, SLAC-PUB-17699, BONN-TH-2022-22",
    doi = "10.1007/JHEP12(2022)066",
    journal = "JHEP",
    volume = "12",
    pages = "066",
    year = "2022"
}

@article{Daleo:2006xa,
    author = "Daleo, A. and Gehrmann, T. and Maitre, D.",
    title = "{Antenna subtraction with hadronic initial states}",
    eprint = "hep-ph/0612257",
    archivePrefix = "arXiv",
    reportNumber = "ZU-TH-24-06",
    doi = "10.1088/1126-6708/2007/04/016",
    journal = "JHEP",
    volume = "04",
    pages = "016",
    year = "2007"
}

\end{document}